\documentclass[12pt]{iopart}
\usepackage{epsf}
\newcommand{\be}{\begin{equation}}
\newcommand{\ee}{\end{equation}}
\newcommand{\bea}{\begin{eqnarray}}
\newcommand{\eea}{\end{eqnarray}}

\usepackage{epsfig}
\usepackage{graphicx}
\usepackage{dcolumn}
\usepackage{bm}
\usepackage{epstopdf}
\usepackage{subfigure}

\begin{document}

\title[The Hubbard model description of the...]
{The Hubbard model description of the photoemission TCNQ singular features}
\author{J. M. P. Carmelo}
\address{GCEP-Center of Physics, University of Minho, Campus Gualtar, P-4710-057
Braga, Portugal} \address{Department of Physics, Massachusetts Institute of
Technology, Cambridge, Massachusetts 02139-4307}
\author{K. Penc}
\address{Research Institute for Solid State Physics and Optics, H-1525 Budapest,
P.O.B. 49, Hungary}
\author{P. D. Sacramento}
\address{Departamento de F\'{\i}sica and CFIF, Instituto Superior T\'ecnico,
P-1049-001 Lisboa, Portugal}
\author{M. Sing and R. Claessen}
\address{Experimentelle Physik 4, Universit\"at W\"urzburg Am Hubland, D-97074
W\"urzburg, Germany}


\begin{abstract}
In this paper we use the pseudofermion dynamical theory (PDT) in the study of the
one-electron removal singular spectral features the one-dimensional Hubbard model. The
PDT reveals that in the whole $(k,\,\omega)$-plane such features are of power-law type
and correspond to well defined lines of three types: the charge singular branch lines,
the spin singular branch lines, and the border lines. One of our goals is the study of
the momentum and energy dependence of the spectral-weight distribution in the vicinity of
such lines. We find that the charge and spin branch lines correspond to the main
tetracyanoquinodimethane (TCNQ) peak dispersions observed by angle-resolved photoelectron
spectroscopy in the quasi-1D organic conductor
tetrathiafulvalene-tetracyanoquinodimethane (TTF-TCNQ). Our expressions refer to all
values of the electronic density and on-site repulsion $U$. The weight distribution in
the vicinity of the singular spectral lines is fully controlled by the overall
pseudofermion phase shifts. Moreover, the shape of these lines is determined by the
bare-momentum dependence of the pseudofermion energy dispersions.
\end{abstract}


\maketitle

\section{INTRODUCTION }

Early studies of quasi-one-dimensional (1D) compounds have focused on the various
low-energy phases, which are not metallic and correspond to broken-symmetry states
\cite{Kagoshima,Basista,BS}. Recently, the resolution of photoemission experiments has
improved, and the {\it normal} state of these compounds was found to display exotic
spectral properties \cite{spectral0,Ralph,Zwick}. The study of the microscopic mechanisms
behind such properties remains until now an interesting open problem. Indeed, the
finite-energy spectral dispersions recently observed in such metals by angle-resolved
photoelectron spectroscopy (ARPES) reveal significant discrepancies from the conventional
band-structure description \cite{,Kagoshima,spectral0,Ralph,Zwick}.

There is some evidence that the 1D Hubbard model \cite{Lieb,Takahashi} describes
successfully the transport properties and other exotic properties observed in some
low-dimensional materials \cite{properties} and that the electronic correlation effects
described by the model could contain the finite-energy microscopic mechanisms
\cite{spectral0,Ralph} that control the above finite-energy spectral properties. Until
recently very little was known about the finite-energy spectral properties of that model
for finite values of the on-site repulsion $U$. This is in contrast to simpler models
\cite{Lee}. Indeed, usual techniques such as bosonization \cite{Schulz} and
conformal-field theory \cite{Woy,Ogata,Kawakami,Frahm,Brech,CFT,Ogata91,Karlo,algebras}
do not apply at finite energy. Valuable qualitative information can be obtained for
$U\rightarrow\infty$ by use of the method of Refs. \cite{Penc96,Penc97}. However,
a quantitative description of the finite-energy spectral properties of quasi-1D metals
requires the solution of the problem for finite values of the on-site Coulombian
repulsion $U$. The method of Ref. \cite{Sorella} refers to features of the insulator
phase. For $U\approx 4t$, where $t$ is the transfer integral, there are numerical results
for the one-electron spectral function \cite{Senechal}. Unfortunately, the latter results
provide very little information about the microscopic mechanisms behind the finite-energy
spectral properties.

Recently, the preliminary use of the finite-energy holon and spinon representation
introduced in Ref. \cite{I} and of the related pseudofermion description of Refs.
\cite{V-1,S0}, revealed that most singular features of the one-electron removal spectral
function correspond to separate charge and spin branch lines \cite{spectral0,Ralph}.
Interestingly, the singular branch lines associated with the one-electron removal
spectral function show quantitative agreement with the TCNQ peak dispersions observed by
ARPES in the quasi-1D organic conductor TTF-TCNQ \cite{spectral0,Ralph,Zwick}. However,
these studies provide no information about the momentum and energy dependence of the
spectral-weight distribution in the vicinity of the charge and spin branch lines. A
preliminary study of that dependence was recently presented in short form in Ref.
\cite{EPL}. Shortly after a study of the same problem by means of the dynamical density
matrix renormalization group (DDMRG) method led to very similar results \cite{Eric}.

The main goal of this paper is the extension of the preliminary results presented in Ref.
\cite{EPL}. Our investigation relies on the pseudofermion dynamical theory (PDT)
introduced in Ref. \cite{V-1}. Such a theory provides the general finite-energy
spectral-weight distributions for the metallic phase of the model (\ref{H}) for all
values of energy and momentum. However, the amount of one-electron removal weight in
regions away from the singular spectral lines is small. Therefore, here we limit our
studies to the vicinity of such lines. Our investigation focus on the one-electron
removal spectral function for all values of $U$ and electronic density in the vicinity of
the singular charge and spin branch lines, which turn out to be the most important
spectral features for the description of the unusual TCNQ photoemission spectral lines.
The evaluation of the small spectral-weight distributions away from the singular spectral
features considered here requires the use of involved numerical calculations which will be
carried out elsewhere.

The pseudofermion description refers to the {\it pseudofermion subspace} (PS) \cite{V-1}.
The one-electron removal excitations studied in this paper are contained in such a
Hilbert subspace. The PDT refers to all energy scales but in the limit of low energy leads 
to the known conformal-field-theory spectral-function and correlated-function expressions \cite{LE}. 
In Ref. \cite{super} the PDT is combined with the
renormalization group for the study of the microscopic mechanisms behind the phase
diagram observed in the (TMTTF)$_2$X and (TMTSF)$_2$X series of quasi-1D organic
compounds. These studies are consistent with the latter phase diagram and explain why
there are no superconducting phases in TTF-TCNQ. The PDT used in this paper in the
study of the one-electron removal spectral function of the 1D Hubbard model applies to 
related integrable interacting problems \cite{tj} and therefore has wide applicability.

The paper is organized as follows: In Sec. I we introduce the one-electron removal
spectral-function problem, the 1D Hubbard model, and some basic information about the
pseudofermion description. The introduction to the PDT and of the corresponding general
spectral-function expressions used in our investigation is the subject of Sec. II.
Section III is devoted to the the study of the weight distributions for one-electron
removal. Moreover, we investigate the limiting behavior of the spectral function in the
vicinity of the branch lines both for $U/t\rightarrow 0$ and $U/t\rightarrow\infty$. The
discussion of the relation of our theoretical predictions to the TCNQ branch lines
observed by angle-resolved photoelectron spectroscopy in the quasi-1D organic compound
TTF-TCNQ and the concluding remarks are presented in Sec. IV.

\subsection{THE MODEL AND THE ONE-ELECTRON REMOVAL SPECTRAL FUNCTIONS}

The 1D Hubbard model reads,
\begin{equation}
\hat{H}=-t\sum_{j,\,\sigma}[c_{j,\,\sigma}^{\dag} c_{j+1,\,\sigma} + h.
c.]+U\sum_{j}\hat{n}_{j,\uparrow}\hat{n}_{j,\downarrow} \, , \label{H}
\end{equation}
where $c_{j,\,\sigma}^{\dagger}$ and $c_{j,\,\sigma}$ are spin-projection $\sigma
=\uparrow ,\downarrow$ electron operators at site $j=1,2,...,N_a$ and
$\hat{n}_{j,\,\sigma}=c_{j,\,\sigma}^{\dagger}\,c_{j,\,\sigma}$. The model (\ref{H})
describes $N_{\uparrow}$ spin-up electrons and $N_{\downarrow}$ spin-down electrons in a
chain of $N_a$ sites. We denote the electronic number by $N=N_{\uparrow}+N_{\downarrow}$.
The number of lattice sites $N_a$ is even and very large. For simplicity, we use units
such that both the lattice spacing and the Planck constant are one. In these units the
chain length $L$ is such that $L=N_a$. Our results refer to periodic boundary conditions.
We consider an electronic density $n=n_{\uparrow }+n_{\downarrow}$ in the range $0<n< 1$
and a spin density $m=n_{\uparrow}-n_{\downarrow}$ such that $m\rightarrow 0$, where
$n_{\sigma}=N_{\sigma}/L$ and $\sigma =\uparrow ,\,\downarrow$. However, the calculations
are performed for finite values of the spin density $m$. For $U/t>0$ the limit
$m\rightarrow 0$ is taken in the end of the calculation and leads to the correct $m=0$
results. The Fermi momentum is $k_F=\pi n/2$ and the electronic charge reads $-e$.

The one-electron removal spectral function $B (k,\,\omega)$ is given by,
\begin{eqnarray}
B (k,\,\omega) & = & \sum_{\sigma}\sum_{f}\, \vert\langle f\vert\, c_{k,\,\sigma} \vert
GS\rangle\vert^2\,\delta\Bigl( \omega + E_{f} - E_{GS}\Bigr) \, , \hspace{0.5cm} \omega <
0  \, . \label{1p}
\end{eqnarray}
Here $c_{k,\,\sigma}$ is an electron annihilation operator of momentum $k$ and $\vert
GS\rangle$ denotes the initial $N$-electron ground state. The $f$ summation runs over the
$N-1$-electron excited energy eigenstates and $[E_{f}-E_{GS}]$ are the corresponding
excitation energies. We use an extended momentum scheme such that $k\in
(-\infty,\,+\infty)$ for the expression given in Eq. (\ref{1p}), yet it is a simple
exercise to obtain the corresponding spectral function expression for the first Brillouin
zone.

\subsection{THE PSEUDOFERMION DESCRIPTION}

The pseudofermion description \cite{I,V-1,S0,II} involves $c\nu$ pseudofermion branches
such that $\nu =0,1,2,...$ and $s\nu$ pseudofermion branches where $\nu =1,2,...$.
However, most of our study refers to the two branches that contribute to the one-electron
removal dominant processes considered below. For the one-electron removal spectral
function corresponding to densities $0<n<1$ and $m\rightarrow 0$ the above-mentioned
dominant processes involve the $c0$ pseudofermions and $s1$ pseudofermions of Ref.
\cite{V-1}. The pseudofermions are related to the pseudoparticles studied in Refs.
\cite{I,II} by a canonical transformation. Such a transformation introduces shifts of
order $1/L$ in the pseudoparticle discrete momentum values and leaves all other
pseudoparticle properties invariant. As a result of such momentum shifts and in contrast
to the pseudoparticles, the corresponding pseudofermions have no residual-interaction
energy terms.

Most of our final expressions refer to densities $0<n<1$ and $m\rightarrow 0$. According
to the studies of Refs. \cite{I,II,V-1,S0}, the $c0$ pseudofermion is a spin-less and
$\eta$-spin-less object that carries charge $-e$ and the $s1$ pseudofermion is a
charge-less and spin-zero two-spinon composite object. (That charge value corresponds to
the description of the transport of charge in terms of electrons \cite{I}.) An important
point that follows from the above analysis is that while the $c0$ pseudofermion branch is
associated with excitations of charge character, the $s1$ pseudofermion branch describes
spin excitations. The $\alpha\nu =c0,\,s1$ pseudofermions carry canonical-momentum
${\bar{q}} = q + Q^{\Phi}_{\alpha\nu} (q)/L$. Here $q$ is the {\it bare-momentum} and
$Q^{\Phi}_{\alpha\nu} (q)/2$ is the overall scattering phase shift given by \cite{S0},
\begin{equation}
Q^{\Phi}_{\alpha\nu} (q)/2 = \pi\sum_{\alpha'\nu'=c0,\,s1}\,
\sum_{q'}\,\Phi_{\alpha\nu,\,\alpha'\nu'}(q,q')\, \Delta N_{\alpha'\nu'}(q') \, ;
\hspace{1cm} \alpha\nu = c0,\,s1\, . \label{Qcan1j}
\end{equation}
On the right-hand side of this equation $\Delta N_{\alpha\nu}(q)=\Delta
{\cal{N}}_{\alpha\nu} ({\bar{q}})$ is the bare-momentum distribution function deviation
$\Delta N_{\alpha\nu} (q) = N_{\alpha\nu} (q) - N^0_{\alpha\nu} (q)$ and the elementary
two-pseudofermion phase shifts $\Phi_{\alpha\nu,\,\alpha'\nu'}(q,q')$ in units of $\pi$
are such that $+\pi\,\Phi_{\alpha\nu,\,\alpha'\nu'}(q,q')$ (and
$-\pi\,\Phi_{\alpha\nu,\,\alpha'\nu'}(q,q')$) gives the phase shift acquired by the
bare-momentum $q$ $\alpha\nu$ pseudofermion or hole wave function when such an object is
scattered by a bare-momentum $q'$ $\alpha'\nu'$ pseudofermion (and $\alpha'\nu'$
pseudofermion hole) created under a ground-state - excited-energy-eigenstate transition
\cite{S0}. The bare-momentum distribution-function deviations $\Delta
N_{\alpha'\nu'}(q')$ of Eq. (\ref{Qcan1j}) result from such a transition. In expression
(\ref{Qcan1j}) these deviations refer to the $c0$ and $s1$ branches only and
thus describe excited energy eigenstates
generated by one-electron removal dominant processes. The corresponding general
expression is given in Eq. (2) of Ref. \cite{S0} and involves summations over all
pseudofermion branches.

Note that for each $\alpha\nu$ branch the continuum bare-momentum $q$ corresponds to a
set of discrete bare-momentum values $q_j$ such that $q_{j+1}-q_j =2\pi/L$. Here
$j=1,2,...,N^*_{\alpha\nu}$ and the number
$N^*_{\alpha\nu}=N_{\alpha\nu}+N_{\alpha\nu}^h$ is given in Eqs. (B6)-(B8) and (B11) of
Ref. \cite{I}. $N_{\alpha\nu}$ and $N_{\alpha\nu}^h$ denote the number of $\alpha\nu$
pseudofermions and $\alpha\nu$ pseudofermion holes, respectively. $N^*_{\alpha\nu}$
equals the number of sites of the effective $\alpha\nu$ lattice \cite{S0}, which plays an
important role in the pseudoparticle and pseudofermion descriptions. For the $\alpha\nu
=c0,\,s1$ branches the number $N_{\alpha\nu}^h$ reads,
\begin{equation}
N^h_{c0}=[N_a -N_{c0}] \, ; \hspace{0.5cm} N^h_{s1} = N_{c0} - 2\sum_{\nu'=1}^{\infty}
N_{s\nu'} \, , \label{Nhs1}
\end{equation}
where in our case $N^h_{s1} = N_{c0} - 2N_{s1}$. We used Eq. (B.11) of Ref. \cite{I} to
derive the expression given in Eq. (\ref{Nhs1}) for $N^h_{s1}$, such that $N^h_{s1}=0$
for the $m\rightarrow 0$ initial ground state.

Although the $\alpha\nu$ pseudoparticles carry bare-momentum $q$ \cite{I,II}, one can
also label the corresponding $\alpha\nu$ pseudofermions by that bare-momentum. Indeed,
the latter pseudofermions carry canonical-momentum ${\bar{q}} = q + Q^{\Phi}_{\alpha\nu}
(q)/L$, but this latter expression provides an one-to-one relation between the
bare-momentum $q$ and the canonical-momentum ${\bar{q}}$. The pseudoparticles have
residual-interaction energy terms which do not allow the expression of the electronic
spectral functions as a convolution of $\alpha\nu$ pseudoparticle spectral functions
\cite{V-1}. A property which plays a central role in the PDT is that for the
corresponding $\alpha\nu$ pseudofermions, such a residual-interaction energy terms are
exactly canceled by the overall scattering phase shift $Q^{\Phi}_{\alpha\nu}(q)/2$. By
canceling the residual interactions exactly, the associated canonical-momentum shift
$Q^{\Phi}_{\alpha\nu}(q)/L$ transfers the information recorded in these interactions over
to the pseudofermion canonical-momentum. It is found in Ref. \cite{S0} that the
$\alpha\nu$ pseudofermion or $\alpha\nu$ pseudofermion hole overall phase shift acquired
under a ground-state - excited-energy-eigenstate transition is given by
$Q_{\alpha\nu}(q)/2=Q^0_{\alpha\nu}/2+Q^{\Phi}_{\alpha\nu}(q)/2$. Here
$Q^{\Phi}_{\alpha\nu}(q)/2$ is provided in Eq. (\ref{Qcan1j}) and
$Q^0_{\alpha\nu}/2=0,\,\pm\pi/2$ is a $\alpha\nu$ pseudoparticle or hole overall
scattering-less phase shift whose value is well defined for each excitation subspace
spanned by energy eigenstates with the same pseudofermion numbers \cite{V-1,S0}. It is
such that under a ground-state - excited-energy-eigenstate transition the $\alpha\nu$
pseudoparticle and hole discrete bare-momentum value $q_j$ is shifted by
$Q^0_{\alpha\nu}(q_j)/L$. The overall phase shift $Q_{\alpha\nu}(q)/2$ leads to a
corresponding canonical-momentum shift $Q_{\alpha\nu}(q_j)/L$ for the discrete
canonical-momentum values of the $\alpha\nu$ pseudofermions and holes. According to the
studies of Ref. \cite{S0}, the overall phase shift $Q_{\alpha\nu}(q)/2$ is associated
with the $\alpha\nu$ pseudofermion or hole $S$ matrix $\exp{\{i Q_{\alpha\nu}(q)\}}$.

The pseudofermions have energy bands $\epsilon_{c0} (q)$ and $\epsilon_{s1} (q)$ such
that $\vert\,q\vert\leq\pi$ and $\vert\,q\vert\leq k_F$, respectively. These energy
dispersions are plotted in Figs. 6 and 7 of Ref. \cite{II} and are defined by Eqs. (C.15)
and (C.16) of Ref. \cite{I}, respectively. Also the group velocity $v_{\alpha\nu}(q) =
\partial\epsilon_{\alpha\nu}(q)/\partial q$
and the {\it Fermi-point} velocity $v_{\alpha\nu}\equiv v_{\alpha\nu}(q^0_{F\alpha\nu})$
appear in the spectral-function expressions. Here $q^0_{Fc0}=2k_F$ and
$q^0_{Fs1}=k_{F\downarrow}$ define the ground-state {\it Fermi points} \cite{I,V-1}. In
Sec. III we confirm that the pseudofermion energy bands fully determine the shape of the
one-electron removal spectral-function in the proximity of the branch lines studied in
that section. In the ground state the $s1$ pseudofermion band is filled and the $c0$
pseudofermions occupy the bare-momentum domain $0\leq\vert\,q\vert\leq 2k_F$ (leaving
$2k_F<\vert\,q\vert\leq\pi$ empty).

\section{THE GENERAL SPECTRAL-FUNCTION EXPRESSIONS USED IN OUR STUDY}

The one-electron removal problem studied in this paper involves only the PDT dominant
processes defined in Ref. \cite{V-1}. The dominant (and non-dominant) processes
correspond to the $i=0$ (and $i>0$) terms on the right-hand side of the general
spectral-function expression given in Eq. (41) of 
the first paper of that reference. Furthermore, only such
dominant processes contribute to the one-electron removal spectral-function power-law
expressions obtained in this paper. The initial ground state and excited energy
eigenstates of Eq. (\ref{1p}) that are associated with the one-electron removal dominant
processes can be expressed in terms of occupancy configurations of the $c0$ and $s1$
pseudofermions. The first step of the evaluation of the spectral function (\ref{1p})
involves the expression of the electronic operators in terms of these quantum objects, as
described in Ref. \cite{V-1}. For one-electron removal, the dominant processes involve
creation of one pseudofermion hole both in the bands $\epsilon_{c0} (q)$ and
$\epsilon_{s1} (q)$. The excited energy eigenstates generated by these one-electron
removal dominant processes belong to subspaces whose $c0$ pseudofermion and $s1$
pseudofermion hole number ground-state deviations are given by,
\begin{equation}
\Delta N^h_{c0}=-\Delta N_{c0}=1\, ; \hspace{0.5cm}\Delta N_{s1}^h=1 \, .
\label{0-CPHS-LHB-c-}
\end{equation}
Moreover, as discussed in Secs. III and IV, the main one-electron removal
spectral-function singular features are associated with the charge $c0$ and spin $s1$
pseudofermion branch lines.

The domain of the $(k,\,\omega)$-plane whose spectral weight is generated by one-electron
removal dominant processes is contained in the region of negative $\omega/t$ values of
Fig. 1 of Ref. \cite{EPL}. (That figure uses the extended momentum scheme also used
here.) Such a domain is limited above by the $s$ line for momentum values between $k=0$
and $k=k_F$, $c''$ line from $k=k_F$ until that line reaches the $c'$ line at $k=2k_F$,
$c'$ line from the momentum $k=2k_F$ until $k=3k_F$, and $s$ line between the momentum
values $k=3k_F$ and $k=5k_F$. The same domain is limited below by the lowest line of the
figure.

Use of the general PDT distributions studied in Ref. \cite{V-1} for the one-electron
removal problem involves all dominant processes associated with the pseudofermion-number
deviations given in Eq. (\ref{0-CPHS-LHB-c-}). In addition to small-momentum and
low-energy $c0$ and $s1$ pseudofermion particle-hole processes, which conserve the
pseudofermion numbers, such dominant processes involve creation of one $c0$ pseudofermion
hole and one $s1$ pseudofermion hole at bare-momentum values $q$ and $q'$, respectively.
When both the bare-momentum values $q$ and $q'$ of the two created pseudofermion holes
are away from the {\it Fermi points} and such that $v_{c0}(q) \neq v_{s1}(q')$, the
corresponding dominant processes do not lead to singular spectral features and generate
the one-electron removal spectral weight for $(k,\,\omega)$-plane regions away from these
features. Since the latter weight is small, in this paper we do not study its intensity
distribution in the $(k,\,\omega)$-plane. A first type of singular feature corresponds to
lines generated by such processes where both created objects move with the same group
velocity, $v_{c0}(q) =v_{s1}(q')$, and the spectral feature corresponds to a border line,
$\omega =\omega_{BL} (k)=[\pm\epsilon_{c0}(q)-\epsilon_{1s}(q')]\,\delta_{v_{c0}(q)
,\,v_{s1}(q')}$, in the $(k,\,\omega)$-plane. In this case the spectral function reads,

\begin{equation}
B (k,\,\omega)\approx C_{BL} (k)\,(\omega -\omega_{BL} (k))^{-{1/2}} \, ,
\label{B-border}
\end{equation}
in the vicinity and just above such a line. However, as discussed in Sec. IV, the only
existing border line for the density of TTF-TCNQ leads to a weak TCNQ spectral
feature mentioned in Sec. IV, which we do not study in Sec. III.

The second type of spectral feature corresponds to the branch lines studied in this
paper, such that either $q$ or $q'$ equals one of the corresponding pseudofermion branch
{\it Fermi points}. Thus, such features are generated by processes where a $\alpha\nu$
pseudofermion hole is created for all the available values of bare-momentum $q$ and a
second $\alpha'\nu'$ pseudofermion hole is created at one of its two {\it Fermi points}
$\pm\,q^0_{F\alpha'\nu'}$, where $\alpha\nu =c0,\,s1$ and $\alpha'\nu' =s1,\,c0$,
respectively. According to the general theory of Ref. \cite{V-1}, for densities $0<n<1$
and $0<m<n$ the functional,
\begin{equation}
2\Delta_{\alpha\nu}^{\iota} = \Bigl(\iota\,\Delta N_{\alpha\nu,\,\iota}^{0,F}+
{Q_{\alpha\nu} (\iota\,q^0_{F\alpha\nu})\over 2\pi}\Bigr)^2 = \Bigl(\iota\,\Delta
N_{\alpha\nu,\,\iota}^F+ {Q^{\Phi}_{\alpha\nu} (\iota\,q^0_{F\alpha\nu})\over
2\pi}\Bigr)^2 \, , \label{Delta}
\end{equation}
where $\iota =\pm 1$ and $\alpha\nu = c0,\,s1$
controls the exponents of the spectral-function power-law expressions in the vicinity of
the charge and spin branch lines. Here $\Delta N_{\alpha\nu,\,\iota}^F$ is the deviation
in the number of $\alpha\nu =c0,\,s1$ pseudofermions at the {\it Fermi points} and
$Q^{\Phi} (\iota\,q^0_{F\alpha\nu})/2$ and $Q_{\alpha\nu}
(\iota\,q^0_{F\alpha\nu})/2=Q^0_{\alpha\nu}/2+ Q^{\Phi} (\iota\,q^0_{F\alpha\nu})/2$ are
the overall scattering and overall, respectively, phase shifts of $\alpha\nu$
pseudofermion or hole scatterers at the bare-momentum {\it Fermi values}
$\iota\,q^0_{F\alpha\nu}=\pm\,q^0_{F\alpha\nu}$. Note that the deviation $\Delta
N_{\alpha\nu,\,\iota}^F =\iota\,\Delta N_{\alpha\nu,\,\iota}^{0,F} +
Q^0_{\alpha\nu}/2\pi$ involves both contributions from the scattering-less overall phase
shift $Q^0_{\alpha\nu}/2$ and the number deviation $\Delta N_{\alpha\nu,\,\iota}^{0,F}$
generated by the creation or annihilation pseudofermion processes at the {\it Fermi
points}. We also consider the  $\alpha\nu =c0,\,s1$ current number deviation $2\Delta
J^F_{\alpha\nu}=\Delta N^F_{\alpha\nu ,\,+1}-\Delta N^F_{\alpha\nu ,\,-1}$.

For excited energy eigenstates generated by processes involving pseudofermion occupancies
in the vicinity of the $c0$ or $s1$ {\it Fermi points}, the functional (\ref{Delta}) and
the corresponding branch-line exponent expressions involve the following {\it Fermi}-point
two-pseudofermion phase-shift parameters,
\begin{equation}
\xi^j_{\alpha\nu\,\alpha'\nu'}= \delta_{\alpha,\,\alpha'}\,\delta_{\nu,\,\nu'} +
\sum_{\iota=\pm 1}(\iota^j)\,\Phi_{\alpha\nu,\,\alpha'\nu'}
(q^0_{F\alpha\nu},\iota\,q^0_{F\alpha'\nu'}) \, ; \hspace{0.5cm} j=0,\,1 \, ,
\label{xi}
\end{equation}
where $\alpha\nu = c0,\, s1$ and $\alpha'\nu' = c0,\, s1$.
In the limit $m\rightarrow 0$, these parameters are given by $\xi^0_{c0\,c0}=1/\xi_0$,
$\xi^0_{c0\,s1}=0$, $\xi^0_{s1\,c0}=-1/\sqrt{2}$, $\xi^0_{s1\,s1}=\sqrt{2}$,
$\xi^1_{c0\,c0}=\xi_0$, $\xi^1_{c0\,s1}=\xi_0/2$, $\xi^1_{s1\,c0}=0$, and
$\xi^1_{s1\,s1}=1/\sqrt{2}$. Here $\xi_0$ is the parameter defined in Eq. (74) of Ref.
\cite{92} and in the text above that equation. It is such that $\xi_0\rightarrow\sqrt{2}$
and $\xi_0\rightarrow 1$ as $U/t\rightarrow 0$ and $U/t\rightarrow\infty$, respectively.

We emphasize that the limits $m\rightarrow 0$, $U/t\rightarrow 0$ and $U/t\rightarrow 0$,
$m\rightarrow 0$ do not commute and lead to different values for the parameters
(\ref{xi}): while for $m\rightarrow 0$, $U/t\rightarrow 0$ one finds
$\xi^0_{c0\,c0}\rightarrow 1/\sqrt{2}$, $\xi^0_{c0\,s1}\rightarrow 0$,
$\xi^0_{s1\,c0}\rightarrow -1/\sqrt{2}$, $\xi^0_{s1\,s1}\rightarrow \sqrt{2}$,
$\xi^1_{c0\,c0}\rightarrow\sqrt{2}$, $\xi^1_{c0\,s1}\rightarrow 1/\sqrt{2}$,
$\xi^1_{s1\,c0}\rightarrow 0$, and $\xi^1_{s1\,s1}\rightarrow 1/\sqrt{2}$, for
$U/t\rightarrow 0$, $m\rightarrow 0$ the result is $\xi^j_{c0\,c0}\rightarrow 1$,
$\xi^j_{c0\,s1}\rightarrow 0$, $\xi^j_{s1\,c0}\rightarrow 0$, and
$\xi^j_{s1\,s1}\rightarrow 1$ both for $j=0$ and $j=1$. Also the phase shifts
$\Phi_{s1,\,c0}(k_{F\downarrow},\,q)$ and $\Phi_{s1,\,s1}(k_{F\downarrow},\,q)$ have
different values in the limits $m\rightarrow 0$, $U/t\rightarrow 0$ and $U/t\rightarrow
0$, $m\rightarrow 0$, respectively. Therefore, for some excitations the spin functionals
$2\Delta_{s1}^{-1}$ and $2\Delta_{s1}^{+1}$ have different values in these two limits.
For such excitations we provide below the spectral-function exponent associated with the
limit $U/t\rightarrow 0$, $m\rightarrow 0$, which is that which leads to the correct
$U/t=0$ spectral-function behavior for $m=0$. This can be confirmed by studying the limit
of the corresponding exponents as $U/t\rightarrow 0$ for $m>0$.

\subsection{GENERAL EXPRESSIONS IN THE VICINITY OF THE PSEUDOFERMION BRANCH
LINES}

Here and in section III we label the $\alpha\nu$ pseudofermions by their bare-momentum
$q$, which is related to the canonical-momentum as ${\bar{q}} = q + Q^{\Phi}_{\alpha\nu}
(q)/L$. The momentum values $k$ of the one-electron removal $\alpha\nu$ branch lines
points $(k,\,l\omega_{\alpha\nu}(k))$ are determined through the bare-momentum value $q$
of the created $\alpha\nu$ pseudofermion or hole by the following parametric equations,
\begin{eqnarray}
k & = & -\Bigl[k_0 -q\Bigr] \, ; \hspace{1cm} q = \Bigl[k+k_0\Bigr] \, ; \hspace{1cm}
\omega_{\alpha\nu} =
-\epsilon_{\alpha\nu} (q) \, ; \nonumber \\
k_0 & = & 4k_F\,\Delta J^F_{c0} + 2k_{F\downarrow}\,\Delta J^F_{s1} \, . \label{*line}
\end{eqnarray}

According to the studies of Ref. \cite{V-1}, the one-electron removal spectral function
(\ref{1p}) has in the vicinity of the $\alpha\nu$ branch lines the following power-law
expression for {\it finite} values of the energy $\omega <0$ such that $-(\omega
+\omega_{\alpha\nu}(q))$ is small and positive and $\omega_{\alpha\nu}(q) \neq 0$ where
the point ($k,\,-\omega_{\alpha\nu}(q)$) belongs to the branch line,
\begin{eqnarray}
B (k,\,\omega) & \approx & C_{\alpha\nu} (q)\Bigl({-[\omega +\omega_{\alpha\nu}(q)]\over
4\pi\sqrt{v_{c0}\,v_{s1}}}\Bigr)^{\zeta_{\alpha\nu} (q)} \, ; \hspace{0.5cm}
\zeta_{\alpha\nu} (q) > -1 \nonumber
\\
& = & \delta \Bigl(\omega +\,\omega_{\alpha\nu} (q)\Bigr) \, ; \hspace{0.5cm}
\zeta_{\alpha\nu} (q) = -1 \, ; \hspace{0.5cm} \alpha\nu = c0,\, s1 \, . \label{Ichiun}
\end{eqnarray}
Here the first expression is the leading-order term of a power-law expansion in the small
energy deviation $-[\omega +\omega_{\alpha\nu}(k)]$ relative to the branch-line energy
whose exponent $\zeta_{\alpha\nu}(k)$ and pre-factor $C_{\alpha\nu} (k)$ have the
following general form,
\begin{eqnarray}
\zeta_{\alpha\nu} (q) & = & -1 + \zeta_0(q) \, ; \nonumber \\ 
\zeta_0(q) & = & 2\Delta_{c0}^{+1}(q)+2\Delta_{c0}^{-1}(q)
+2\Delta_{s1}^{+1}(q)+2\Delta_{s1}^{-1}(q) \geq 0 \, ; \nonumber \\
C_{\alpha\nu} (q) & = & {{\rm sgn} (q)\,1\over 2\pi}\int_{-{{\rm sgn} (q)\,1\over
v_{s1}}}^{{I_{\alpha\nu}(q)\over v_{\alpha\nu} (q)}}dz\,{F_0 (z)\over [1
-z\,v_{\alpha\nu}(q)]^{\zeta_0(q)}} \geq 0 \, ; \hspace{0.25cm} v_{s1}\leq 
{\vert v_{\alpha\nu}(q)\vert\over I_{\alpha\nu}(q)} \nonumber \\
& = & {{\rm sgn} (q)\,1\over 2\pi}\int_{-{{\rm sgn} (q)\,1\over v_{s1}}}^{{{\rm sgn}
(q)\,1\over v_{s1}}}dz\,{F_0 (z)\over [1 -z\,v_{\alpha\nu}(q)]^{\zeta_0(q)}} \geq 0 \, ;
\hspace{0.25cm} v_{s1}\geq {\vert v_{\alpha\nu}(q)\vert\over I_{\alpha\nu}(q)} \, . \label{zeta*}
\end{eqnarray}
In these expressions $I_{\alpha\nu}(q)=[1+[\omega
+\omega_{\alpha\nu}(q)]/\Omega]$,
the small positive energy $\Omega$ corresponds to the energy range
of the small-momentum and low-energy $c0$ and $s1$ pseudofermion particle-hole elementary
processes, $F_0 (z)$ is an even function of $z$ which is defined in Ref. \cite{V-1}, and
the functional $2\Delta_{\alpha\nu}^{\iota}$ is given in Eq. (\ref{Delta}). (In the
numerical evaluation of the spectral-function expressions the value of $\Omega$ is
determined by the normalization procedure associated with imposing 
the $k$ and $\omega$ spectral-function sum rules \cite{V-1};
The expression provided in Ref. \cite{V-1} for $F_0 (z)$ is valid when
the four parameters $2\Delta_{\alpha\nu}^{\iota}$ are finite;
An expression for $F_0 (z)$ valid when some of these parameters vanish will
be given elsewhere.) A full
quantitative study of the pre-factor $C_{\alpha\nu} (q)$ whose general expression is
provided in Eq. (\ref{zeta*}) involves the numerical derivation of the lowest-peak weight
functional introduced in Ref. \cite{V-1}, which is included in the expression of the
function $F_0 (z)$. Such a quantitative study, which requires involved numerical
calculations, is beyond the goals of this paper and will be carried out elsewhere.
However, the general $C_{\alpha\nu} (q)$ expression of Eq. (\ref{zeta*}) is useful for
our studies, once it can be used to extract information about the behavior of the
pre-factor $C_{\alpha\nu} (q)$ as $U/t\rightarrow 0$ and find out whether for finite
values of $U/t$ that function is vanishing or finite and also what its relative value for
different branch lines is.

A $\alpha\nu$ branch line whose exponent $\zeta_{\alpha\nu} (q)$ is negative for a given
domain of $k = -[k_0 -q]$ values is called a singular branch line. In this case the
weight distribution shows a singular behavior at the branch line, and we expect that the
spectral peaks will be observed in a real experiment. This was confirmed for the present
case of one-electron removal, as discussed in Sec. V. On the other hand, when for a
$(k,\,\omega)$-plane region in the vicinity of the branch line and contained inside the
one-electron removal dominant-weight domain of Fig. 1 of Ref. \cite{EPL} the exponent
(\ref{zeta*}) is such that $0<\zeta_{\alpha\nu} (q)<1$, the spectral feature refers to an
edge branch line. Finally, $0<\zeta_{\alpha\nu} (q)<1$ for regions away from that domain
and $\zeta_{\alpha\nu} (q)>1$ for any $(k,\,\omega)$-plane region are in general a sign
of near absence of spectral weight. For one-electron removal and thus $\omega<0$, the
singular and edge branch lines are represented in Fig. 1 of Ref. \cite{EPL} by solid and
dashed lines, respectively. The dashed-dotted lines of that figure are either limiting
lines for the domain of weight generated by dominant processes or lines associated with
exponents larger than one. The lowest limiting line of the figure corresponds to a
singular but weak spectral feature called a border line, as mentioned in Sec. V.

The general branch-line spectral-function expressions defined by Eqs. (\ref{Ichiun}) and
(\ref{zeta*}) are not valid in the vicinity of the low-energy branch-line end points.
These end points correspond to values of $k$ and $\omega$ such that $\omega\approx
\iota\, v_{\alpha\nu}(k+k_0^F)$ where,
\begin{eqnarray}
k_0^F = k_ 0 - \iota\,q^0_{F\alpha\nu} \, ; \hspace{0.5cm} \alpha\nu = c0,\,s1 \, ,
\hspace{0.25cm} \iota =\pm 1 \, , \nonumber
\end{eqnarray}
and $k_ 0$ is the momentum given in Eq. (\ref{*line}). In this low-energy limit the
physics is that of the so called low-energy Tomanaga-Luttinger liquid (TLL) regime, where
bosonization \cite{Schulz} and conformal-field theory
\cite{Woy,Ogata,Kawakami,Frahm,Brech,CFT,Ogata91,Karlo,algebras} are applicable.

\subsection{LIMITING LOW-ENERGY BEHAVIOR NEAR THE BRANCH-LINE END POINTS}

The $\alpha\nu = c0,\,s1$ branch lines also exist for {\it small} positive values of
$-\omega$. Such a regimen corresponds to the above-mentioned values of $k$ and $\omega$
such that $\omega\approx \iota\, v_{\alpha\nu}(k+k_0^F)$. In this case the expression
(\ref{Ichiun}) does not apply and instead the general PDT leads to the following
one-electron removal spectral-function expression \cite{V-1},
\begin{eqnarray}
B (k,\,\omega) & \approx & C_{\alpha\nu,\,\iota} \Bigl({-[\omega
+\iota\,v_{\alpha\nu}(k+k_0^F)]\over
4\pi\sqrt{v_{c0}\,v_{s1}}}\Bigr)^{-1+\zeta^0_{\alpha\nu,\,\iota}} \, ; \nonumber \\
\zeta^0_{\alpha\nu,\iota} & = & \zeta^0_{\alpha\nu} - 2\Delta_{\alpha\nu}^{-\iota} \geq 0 \,
, \label{zeta*END}
\end{eqnarray}
where $\alpha\nu = c0,\, s1$, $\iota =\pm 1$, and the pre-factor $C_{\alpha\nu,\,\iota}$
is a real and positive number. The low-energy spectral function expression given in Eq.
(\ref{zeta*END}) refers in general to the proximity of a $\alpha\nu =c0,\,s1$ branch-line
end point. In turn, the finite-energy expression (\ref{Ichiun}) applies in the vicinity
of the $\alpha\nu =c0,\,s1$ branch lines when the $\alpha\nu =c0,\,s1$ branch-line group
velocity $v_{\alpha\nu}(q)$ is such that $v_{\alpha\nu}(q)\neq \iota\, v_{\alpha\nu}$. As
the value of $v_{\alpha\nu}(q)$ approaches that of the {\it Fermi point} velocities $\pm
v_{\alpha\nu}$, $v_{\alpha\nu}(q)\rightarrow \iota\, v_{\alpha\nu}$, the spectral
function $k$ and $\omega$ values reach the vicinity of a branch-line end point and thus
the spectral function is given by expression (\ref{zeta*END}) instead of (\ref{Ichiun}).

Interestingly, for $\omega\approx \iota\,v_{\alpha\nu}(k-lk_0^F)$ in the momentum
expression (\ref{*line}), the validity of the spectral-function expression given in Eq.
(\ref{zeta*END}) corresponds to the TLL regime. Therefore, in this limit the above
exponent $-1+\zeta^0_{\alpha\nu,\,\iota}$ provided by the general PDT of Ref. \cite{V-1}
must equal that corresponding to the low-energy universal TLL expressions. Indeed, it is
straightforward to show that the general exponent $-1+\zeta^0_{\alpha\nu,\,\iota}$ of
expression (\ref{zeta*END}) is identical to that given in Eq. (5.7) of Ref. \cite{CFT}.
When applied to specific spectral functions such that $\omega<0$, expression
(\ref{zeta*END}) provides the universal and well known low-energy TLL behavior for the 1D
Hubbard model \cite{Woy,Ogata,Kawakami,Frahm,Brech,CFT,Ogata91,Karlo,algebras},
Tomonaga-Luttinger model \cite{Meden,Voit93,Schon}, and many other models whose
low-energy physics corresponds to the same universality class. When
$-1+\zeta^0_{\alpha\nu,\iota}<0$, such an expression refers to a linear singular spectral
feature.

There is a cross-over region between the finite-energy and low-energy regimens
corresponding to the spectral-function expressions (\ref{Ichiun}) and (\ref{zeta*END}),
respectively. The momentum and energy width corresponding to such a crossover regimen is
very small and is fully controlled by the value of $\vert v_{\alpha\nu}
(q)-\iota\,v_{\alpha\nu}\vert$. The low-energy TLL behavior emerges when $\vert
v_{\alpha\nu} (q)-\iota\,v_{\alpha\nu} \vert \approx \vert
a_{\alpha\nu}(q^0_{F\alpha\nu})(k+k_0^F)\vert$, where $a_{\alpha\nu}(q) = \partial
v_{\alpha\nu}(q)/{\partial q}$ for $\alpha\nu = c0,\, s1$. As the value of the
branch-line bare-momentum $q$ of Eq. (\ref{*line}) approaches $\iota q^0_{F\alpha\nu}$
the behavior (\ref{zeta*END}) is reached. Importantly, for smaller values of $\vert
a_{\alpha\nu} (q^0_{F\alpha\nu})\vert$ the value of $\vert v_{\alpha\nu}
(q)-\iota\,v_{\alpha\nu}\vert $ can remain small for larger values of $\vert
k+k_0^F\vert$ and thus of $\omega\approx \iota\,v_{\alpha\nu}(k+k_0^F)$. It follows that
the momentum and energy widths of the $(k,\omega)$-plane region in the vicinity of the
point $(-k_0^F,\,0)$ where the TLL liquid behavior is valid increase for decreasing
values of $\vert a_{\alpha\nu} (q^0_{F\alpha\nu})\vert$, provided that $v_{\alpha\nu}$ is
finite. For instance, in the limit of zero spin density, $m\rightarrow 0$, the value of
$\vert a_{s1} (q)\vert$ is small in two relatively large $q$ regions in the vicinity of
$q=-k_F$ and $q=+k_F$, respectively, and thus the domain of the corresponding spin $s1$
branch lines where the TLL expression (\ref{zeta*END}) is valid increases in that limit.
(We recall that the general exponent (\ref{zeta*}) cannot be obtained by the TLL
low-energy methods.)

From the use of conformal-field theory \cite{Carmelo97a}, it is known that when the above
branch-line end point $(-k_0^F,\,0)$ is reached through low-energy lines other than the
above branch lines the spectral-function expression is different from (\ref{zeta*END}).
Examples of such points are the points $(k=k_F,\,\omega =0)$ and
$(k=3k_F,\,\omega =0)$ shown in Fig. 1 of Ref. \cite{EPL}. 
The general theory provides spectral-function expressions that
are valid when these points are approached by lines that are contained in the
finite-weight regions and do not cross the $\alpha\nu$ branch lines ending at the same
points. The velocity $v = 1/z = (\omega/(k+k_0^F)$ plays an important role in these
expressions. Indeed, following the studies of Ref. \cite{V-1}, in the proximity of the
point $(-k_0^F,\,0)$ but for values of $k$ and $\omega$ such that $\omega\approx
v\,(k+k_0^F)$, where $v\neq \pm v_{s1},\,\pm v_{c0}$ and depending on the specific point,
the $z=1/v$ domain is bounded by two of the four values $-1/v_{c0}$, $-1/v_{s1}$,
$1/v_{s1}$, and $1/v_{c0}$, the momentum and energy weight-distribution dependence has
the following general expression for finite values of $U/t$,
\begin{equation}
B (k ,\,\omega)\approx {F_0 (1/v)\over 4\pi\,\sqrt{v_{c0}\,v_{s1}}}\,\Bigl({-\omega\over
4\pi\sqrt{v_{c0}\,v_{s1}}} \Bigr)^{-2+\zeta_0} \, ; \hspace{0.5cm} -2+\zeta_0
> -1 \, , \label{Ichiun0}
\end{equation}
where $F_0 (z)$ is the function appearing in Eq. (\ref{zeta*}), which is defined in Ref.
\cite{V-1}, and $\zeta_0$ is the above functional $\zeta_0 =
2\Delta_{c0}^{+1}+2\Delta_{c0}^{-1} +2\Delta_{s1}^{+1}+2\Delta_{s1}^{-1}$. Here
$2\Delta_{\alpha\nu}^{\iota}$ is the functional of Eq. (\ref{Delta}) for the
bare-momentum distribution function deviations associated with the excitations that
control the spectral-weight distribution in the vicinity of the point $(-k_0^F,\,0)$. We
note that the $\omega$ dependence of the spectral-function (\ref{Ichiun0}) is that also
obtained by conformal-field theory in Ref. \cite{Carmelo97a} and by another method in
Ref. \cite{IVa}.

Since the power-law spectral-function expressions (\ref{zeta*END}) and (\ref{Ichiun0})
were studied previously by other methods and refer to the vicinity of low-energy lines
and isolated zero-energy points in the $(k,\,\omega)$-plane, respectively, that are not
of interest for the low-energy phase of the organic compound TTF-TCNQ, here we limit our
study to the more complex problem of the finite-energy branch-line spectral weight.
Indeed, the low-energy phase of TTF-TCNQ is not metallic and refers to broken-symmetry
states \cite{Kagoshima,Basista,BS}, whereas the branch-line spectral features given by
Eq. (\ref{Ichiun}) refer to finite-energy values which correspond to the unusual metallic
state of that organic compound.

\section{THE ONE-ELECTRON REMOVAL BRANCH LINES}

In this section we use the general branch-line expressions and associated quantities
considered above in the study of the one-electron removal finite-energy spectral-function
singular and edge branch lines. The ground-state - excited-energy-eigenstate transitions
to the subspace whose pseudofermion number deviations are given in Eq.
(\ref{0-CPHS-LHB-c-}), generate several $\alpha\nu$ pseudofermion branch lines whose
location in the $(k,\,\omega)$-plane is shown in the Fig. 1 of Ref. \cite{EPL} for
$\omega <0$. The $s\equiv s1$ branch-line shown in the figure, which connects the points
$(k=-k_F,\,\omega =0)$ and $(k=k_F,\,\omega =0)$, is generated by creating the $c0$
pseudofermion hole at one of its {\it Fermi} points, and the $s1$ pseudofermion hole for
bare-momentum values in the domain defined by the inequality $\vert\,q\vert\leq k_F$. We
emphasize that in addition to creation of a $c0$ pseudofermion hole at $q=2k_F$ (and
$q=-2k_F$), this excitation includes a collective bare-momentum shift $Q_{c0}^0/L=+\pi/L$
(and $Q^0_{c0}/L=-\pi/L$) for the whole $c0$ pseudofermion {\it Fermi sea}.

By considering the same processes, plus transferring a $c0$ pseudofermion from the {\it
Fermi point} $-\iota'\,2k_F$ to the {\it Fermi point} $\iota'\,2k_F$, two other $s1$
branch lines are generated, which connect the points $(k=-\iota'\,3k_F,\,\omega =0)$ and
$(k=-\iota'\,5k_F,\,\omega =0)$ where $\iota'=\pm 1$. The $\iota'=-1$ line is labeled by
$s$ in Fig. 1 of Ref. \cite{EPL}, where it appears for $\omega\leq 0$. On the other hand,
there are four $c0$ pseudofermion branches lines which connect the points
$(k=-3k_F,\,\omega =0)$ and $(k=k_F,\,\omega =0)$, $(k=-k_F,\,\omega =0)$ and
$(k=3k_F,\,\omega =0)$, $(k=-5k_F,\,\omega =0)$ and $(k=-k_F,\,\omega =0)$, and
$(k=k_F,\,\omega =0)$ and $(k=5k_F,\,\omega =0)$. The first, second, and fourth of these
lines are labeled by $c$, $c'$, and $c''$, respectively, in Fig. 1 of Ref. \cite{EPL},
where their $k>0$ parts are shown for $\omega\leq 0$. Below we study the spectral-weight
distribution in the vicinity of these seven one-electron removal branch lines.

We start by evaluating the weight distribution corresponding to the first $s1$
pseudofermion branch line mentioned above. The specific form of the general expressions
(\ref{*line}) for the points $(k,\,-\omega_{s1} (k))$ belonging to the $s1$ pseudofermion
branch line in the $m\rightarrow 0$ limit, corresponds to $k_0=0$ and reads,
\begin{equation}
q = k \, ; \hspace{1cm} -\omega_{s1} (q) = \epsilon_{s1} (q) \, . \label{sline}
\end{equation}
Here $\epsilon_{s1} (q)$ is the energy dispersion given in Eq. (C.16) of Ref. \cite{I}
and plotted for $m\rightarrow 0$ in Fig. 7 of Ref. \cite{II}. (The energy dispersions
$\epsilon_{c0} (q)$ and $\epsilon_{s1} (q)$ appearing in other expressions of this
section are those defined in Eqs. (C.15) and (C.16) of Ref. \cite{I} and plotted in Figs.
6 and 7 of Ref. \cite{II}, respectively.) We recall that the $k>0$ part of this $s1$
pseudofermion singular branch line is labeled by $s$ in Fig. 1 of Ref. \cite{EPL}, where
it connects the points $(k=0,\,\omega =\epsilon_{s1} (0))$ and $(k=k_F,\,\omega =0)$. In
this case the general spectral-function expression (\ref{Ichiun}) applies provided that
the specific expression associated with the excitations around the point
$(k,\,-\omega_{s1} (k))$ of the functional $2\Delta_{\alpha\nu}^{\iota}$ defined in Eq.
(\ref{Delta}) is used. This expression is a function of $k=q$ and corresponds to the
$m\rightarrow 0$ limit of the following quantity,
\begin{equation}
2\Delta_{\alpha\nu}^{\iota}(q) = \Bigl\{-\iota\,{\xi^0_{\alpha\nu\,c0}\over 2} -
\Phi_{\alpha\nu,\,s1}(\iota\,q^0_{F\alpha\nu},\,q)\Bigr\}^2  \, ; \hspace{0.5cm}
\alpha\nu = c0, \, s1 \, . \label{Dels}
\end{equation}
Here the value of the two-pseudofermion phase shift $\Phi_{\alpha\nu,\,s1}$ and that of
the two-pseudofermion phase shifts $\Phi_{\alpha\nu,\,\alpha'\nu'}$ appearing in other
expressions of this section is uniquely defined in terms of the solution of a system of
integral equations \cite{V-1} and the parameter $\xi^0_{\alpha\nu\,c0}$ general
expression is provided in Eq. (\ref{xi}).

Direct use of expression (\ref{Ichiun}) in the $m\rightarrow 0$ limit, leads to the
following expression for the one-electron removal spectral function,
\begin{eqnarray}
B (k ,\,\omega) & \approx & C_{s1} (q)\Bigl({-[\omega +\omega_{s1}(q)]\over
4\pi\sqrt{v_{c0}\,v_{s1}}}\Bigr)^{\zeta_{s1} (q)} \, ;
\hspace{0.5cm} \zeta_{s1} (q) > -1 \nonumber \\
& = & \delta \Bigl(\omega+\omega_{s1} (q)\Bigl) \, ; \hspace{0.5cm} \zeta_{s1} (q)= -1 \,
, \label{Bs}
\end{eqnarray}
which corresponds to energy values just below the branch line for $\zeta_{s1} (q) > -1$
and at that line for $\zeta_{s1} (q)= -1$ and to bare-momentum and momentum values in the
range $-k_F<q< k_F$ and $-k_F<k< k_F$, respectively. The pre-factor $C_{s1}(q)$ given in
Eq. (\ref{zeta*}) is finite for all values of the $q$ domain, except in the vicinity of
the branch line end points, where $v_{s1}(q)\approx  \pm v_{s1}$ and provided that
$\zeta_{s1} (q) > -1$ the spectral function is instead of the form given in Eq.
(\ref{zeta*END}). When $\zeta_{s1} (q)= -1$ the second spectral-function expression of
Eq. (\ref{Bs}) applies. It refers to the whole branch-line momentum domain.

As the spin density $m$ approaches zero, we find the following exponent expression valid
for all values of $U/t$ and electronic density $n$,
\begin{eqnarray}
\zeta_{s1}(q) & = & -1 + \sum_{\alpha\nu =c0,\,s1}\,\sum_{\iota =\pm
1}\,\Bigl\{{\xi^0_{\alpha\nu\,c0}\over 2} +
\iota\,\Phi_{\alpha\nu,\,s1}(\iota\,q^0_{F\alpha\nu},\,q)\Bigr\}^2
\nonumber \\
& = & -1 + \sum_{\iota =\pm 1}\Bigl\{{1\over 2\xi_0} +
\iota\,\Phi_{c0,\,s1}(\iota\,2k_F,\,q)\Bigr\}^2 
\nonumber \\
& +  & \sum_{\iota =\pm 1}\Bigl\{-{1\over
2\sqrt{2}} + \iota\,\Phi_{s1,\,s1}(\iota\,k_F,\,q)\Bigr\}^2 \, , \label{zetas}
\end{eqnarray}
where the second expression was obtained by taking the limit $m\rightarrow 0$ in the
first-expression quantities and the parameter $\xi_0$ is defined in Eq. (74) of Ref.
\cite{92}. The dependence of the exponent (\ref{zetas}) on the momentum $k$ is obtained
by combining Eqs. (\ref{sline}) and (\ref{zetas}). The exponent $\zeta_{s1}$ of Eq.
(\ref{zetas}) is negative for all values of momentum and is plotted in Fig. 1 as a
function of the momentum $k$ for $k>0$, several values of $U/t$, and electronic density
$n=0.59$. So the corresponding spectral-function expression (\ref{Bs}) describes a
singular branch line. The exponent $\zeta_{s}$ of the figure is the exponent
(\ref{zetas}) for momentum values $0<k<k_F$, whereas for $k_F<k<3k_F$ $\zeta_{s}$ it
corresponds to an one-electron addition exponent considered in Ref. \cite{EPL}.

While for momentum values $0<k<k_F$ one reaches the same value for the exponent plotted
in Fig. 1 in the limits $m\rightarrow 0$, $U/t\rightarrow 0$ and $U/t\rightarrow 0$,
$m\rightarrow 0$, that value is different for $k_F<k<3k_F$. In this paper we always
consider the limit $U/t\rightarrow 0$, $m\rightarrow 0$, whereas the studies of Ref.
\cite{EPL} considered the limit $m\rightarrow 0$, $U/t\rightarrow 0$. This justifies the
different values of that exponent for $k_F<k<3k_F$ and $U=0$ given in Fig. 1 and in Fig.
2 of Ref. \cite{EPL}, respectively, which otherwise correspond to the same exponent
values. The $U/t\rightarrow 0$ and $U/t\rightarrow\infty$ limiting values of the exponent
(\ref{zetas}) and other exponents obtained below are further discussed at the end of this
section.

\begin{figure}
\includegraphics[width=7cm,height=7cm]{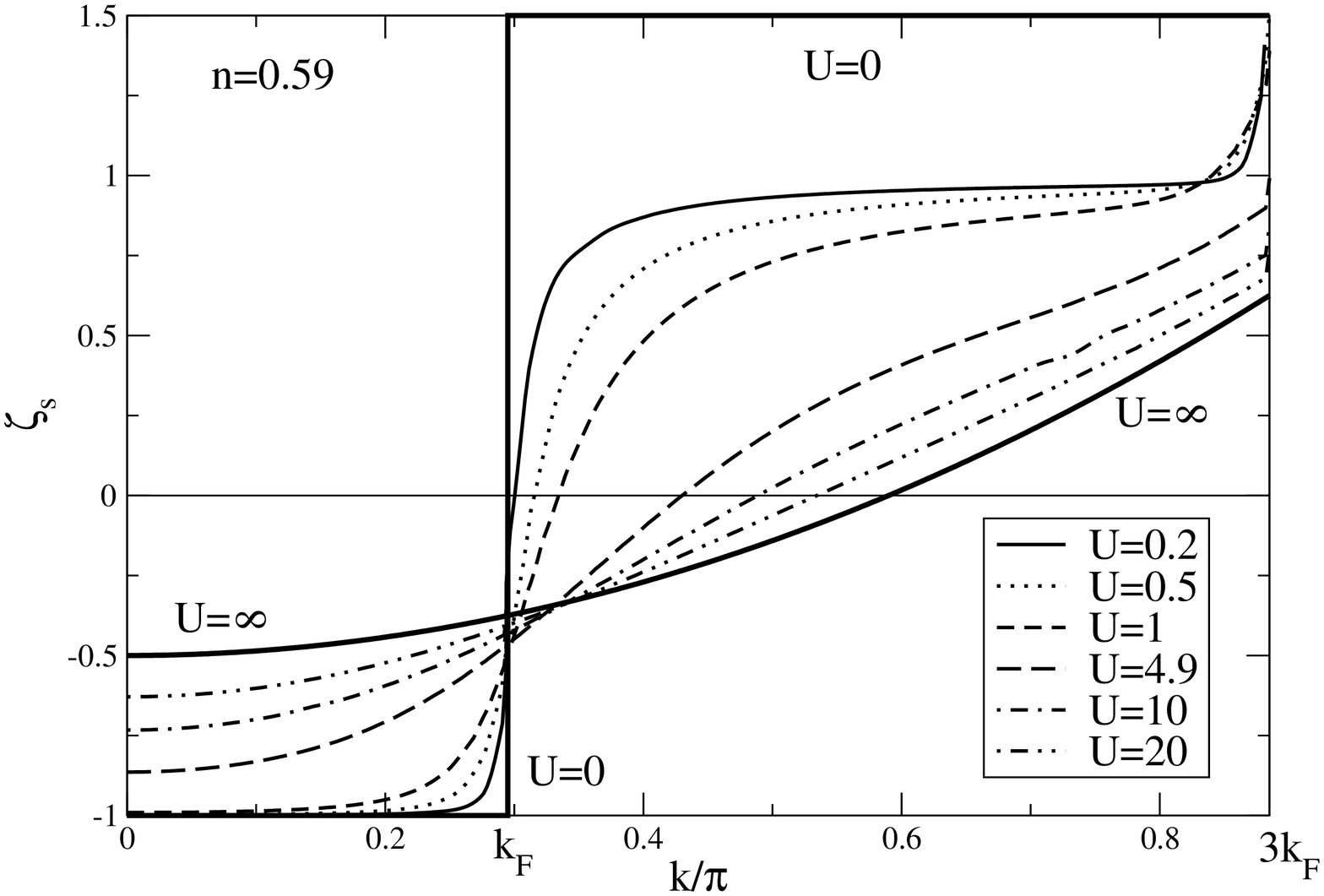}
\caption{\label{fig1} Momentum dependence of the exponents associated with the
one-electron removal spin $s\equiv s1$ branch line of Fig. 1 of Ref. \cite{EPL} for
$0<k<k_F$ and one-electron addition spin branch line of the same figure for $k_F<k<3k_F$.
For the one-electron removal case considered here that exponent is given in Eq.
(\ref{zetas}). (In the figure both these exponents are called $\zeta_{s}$.) We note that
for $U>0$ and in the one-electron removal small momentum domain in the vicinity of the
branch-line end point $k=k_F$, where $v_{s1}(q)\approx v_{s1}$, the exponent plotted here
does not apply, since the spectral function is instead of the form given in Eq.
(\ref{zeta*END}).}
\end{figure}

Similar results are obtained for the other two one-electron removal $s1$ branch lines,
whose exponent was not studied in Ref. \cite{EPL}. We call them $s1,\iota'$ branch lines
where $\iota'=\pm 1$. Here and in the expressions provided below we use the indices
$\iota'=\pm 1$ and $\iota'' =\pm1$ to denote contributions from processes which involve
$c0$ and $s1$ pseudofermions, respectively, created or annihilated at the $m\rightarrow
0$ {\it Fermi points} $\iota'\,2k_F=\pm 2k_F$ and $\iota''\,k_F=\pm k_F$, respectively.
For the $s1,\iota'$ branch lines, the index $\iota'$ refers to a $c0$ pseudofermion
particle-hole process such that a $c0$ pseudofermion is annihilated at $q=-\iota'\,2k_F$
and created at $q=\iota'\,2k_F$, where $\iota'=\pm 1$. The specific form of the general
expressions (\ref{*line}) for the points $(k,\,\omega_{s1,\,\iota'} (k))$ belonging to
the $s1,\iota'$ branch line in the $m\rightarrow 0$ limit, corresponds to $k_0
=\iota'\,4k_F$ and is given by,
\begin{equation}
q = k + \iota'\,4k_F \, ; \hspace{1cm} \omega_{s1,\,\iota'} (q) = -\epsilon_{s1} (q) \, .
\label{slinei}
\end{equation}
The $s1,\,-1$ pseudofermion singular branch line is labeled by $s$ in Fig. 1 of Ref.
\cite{EPL}, where for $\omega/t<0$ it connects the points $(k=3k_F,\,\omega =0)$ and
$(k=5k_F,\,\omega =0)$. In this case the value of the functional (\ref{Delta}) is a
function of $k=q+\iota'\,4k_F$ given by the $m\rightarrow 0$ limit of the parameter,
\begin{equation}
2\Delta_{\alpha\nu}^{\iota}(q) = \Bigl\{-\iota\,{\xi^0_{\alpha\nu\,c0}\over 2}
+\iota'\,\xi^1_{\alpha\nu\,c0} -
\Phi_{\alpha\nu,\,s1}(\iota\,q^0_{F\alpha\nu},\,q)\Bigr\}^2 \, ; \hspace{0.5cm} \alpha\nu
= c0,\,s1 \, . \label{Delsi}
\end{equation}
Use of the general expression (\ref{Ichiun}) in the $m\rightarrow 0$ limit, leads to the
following expression for the one-electron removal spectral function,
\begin{eqnarray}
B (k ,\,\omega) & \approx & C_{s1,\,\iota'} (q)\Bigl({-[\omega +\omega_{s1,\,\iota'}
(q)]\over 4\pi\sqrt{v_{c0}\,v_{s1}}}\Bigr)^{\zeta_{{s1,\,\iota'}} (q)} \, . \label{Bsi}
\end{eqnarray}
This expression corresponds to energy values just below the branch line and to
bare-momentum values in the range $-k_F<q< k_F$ and momentum values in the domains
$-5k_F<k<-3k_F$ and $3k_F<k<5k_F$ for $\iota'=1$ and $\iota'=-1$, respectively. In the
$m\rightarrow 0$ limit, we find the following exponent expression valid for all values of
$U/t$ and electronic density $n$,
\begin{eqnarray}
\zeta_{s,\,\iota'}(q) & = & -1 + \sum_{\alpha\nu =c0,\,s1}\,\sum_{\iota =\pm
1}\,\Bigl\{{\xi^0_{\alpha\nu\,c0}\over 2} -\iota\,\iota'\,\xi_{\alpha\nu,\,c0}^1 +
\iota\,\Phi_{\alpha\nu,\,s1}(\iota\,q^0_{F\alpha\nu},\,q)\Bigr\}^2
\nonumber \\
& = & -1 + \sum_{\iota =\pm 1}\Bigl\{{1\over 2\xi_0} - \iota\,\iota'\,\xi_0 +
\iota\,\Phi_{c0,\,s1}(\iota\,2k_F,\,q)\Bigr\}^2 \nonumber \\
& +  & \sum_{\iota =\pm 1}\Bigl\{-{1\over
2\sqrt{2}} + \iota\,\Phi_{s1,\,s1}(\iota\,k_F,\,q)\Bigr\}^2 \,  . \label{zetasi}
\end{eqnarray}
The dependence of the exponent (\ref{zetasi}) on the momentum $k$ is obtained by
combining Eqs. (\ref{slinei}) and (\ref{zetasi}). For most of the parameter space and
bare-momentum values, this exponent is larger than one and thus the spectral-function
expression (\ref{Bsi}) does not describe a branch line. Consistently, for finite values
of $U/t$ the pre-factor $C_{s1,\,\iota'} (q)$ of Eq. (\ref{Bsi}) has smaller values than
those of the pre-factor $C_{s1}(q)$ appearing in expression (\ref{Bs}). However, for
large values of $U/t$ and bare-momentum values in the vicinity of $\iota'\,k_F$ such an
exponent corresponds to a branch line, as it reaches values smaller than one. We recall
that for small domains in the vicinity of the end points $k=\pm 3k_F$ and $k=\pm 5k_F$
the spectral function is not of the form (\ref{Bsi}), but instead is of the general form
given in Eq. (\ref{zeta*END}).

\begin{figure*}
\includegraphics[width=7cm,height=7cm]{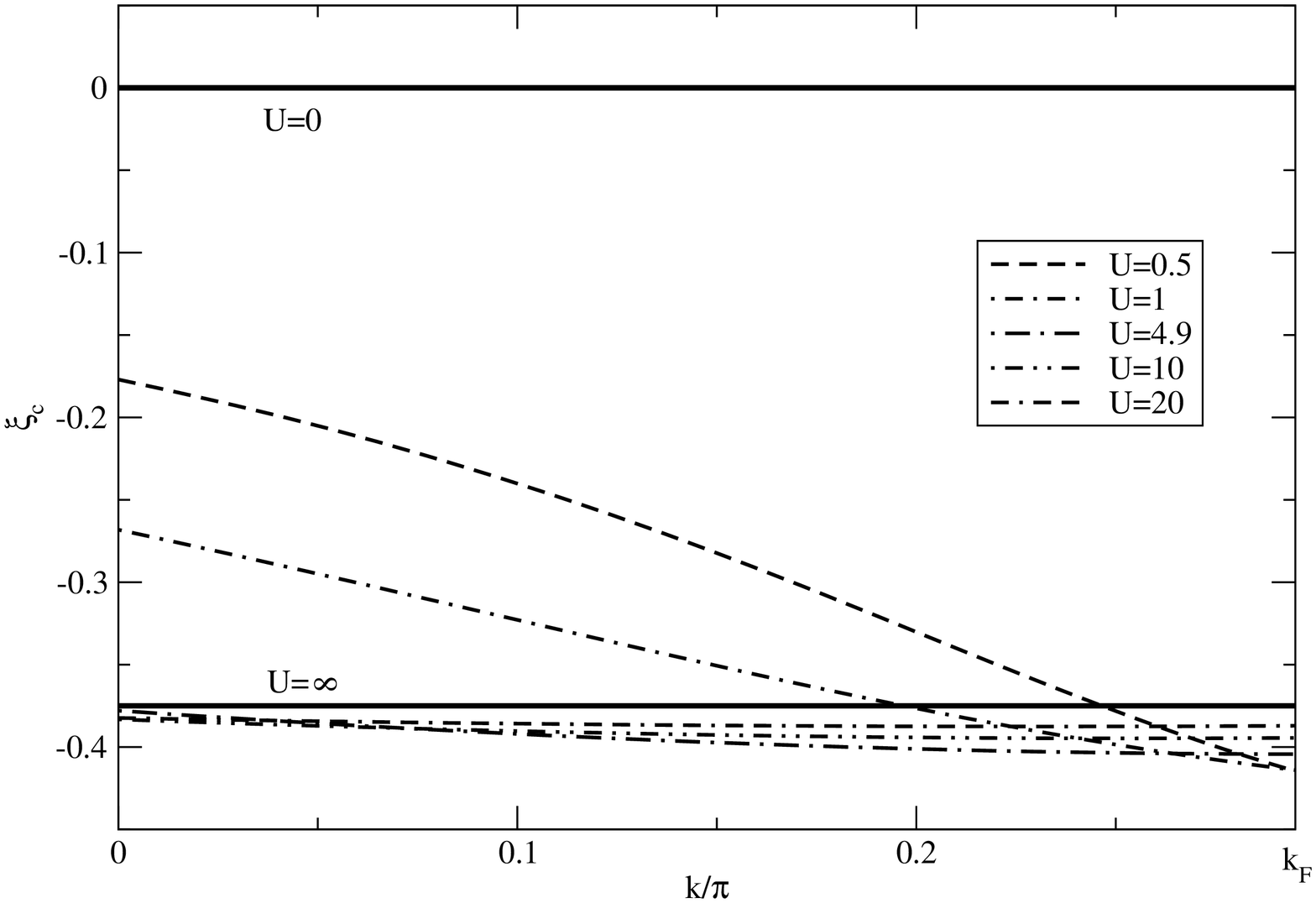}
\caption{\label{fig2} Momentum dependence of the exponent (\ref{zetaci}) along the
one-electron removal $c\equiv c0,\,+1,\,+1$ branch line of Fig. 1 of Ref. \cite{EPL} for
$0<k<k_F$. In the figure that exponent is called $\zeta_c$ and is given by $0$ and $-3/8$
for $U/t\rightarrow 0$ and $U/t\rightarrow\infty$, respectively. We note that for a small
momentum domain in the vicinity of the branch-line end point $k=k_F$, where
$v_{c0}(q)\approx v_{c0}$, this exponent does not apply, since the spectral function is
of the form given in Eq. (\ref{zeta*END}).}
\end{figure*}

Equivalent results are obtained for the four $c0,\iota',\iota''$ branch lines, where
$\iota',\,\iota''=\pm 1$. In this case, the specific form of the general expressions
(\ref{*line}) for the points $(k,\,\omega_{c0,\,\iota',\,\iota''}(k))$ belonging to the
${c0,\,\iota',\,\iota''}$ branch lines in the $m\rightarrow 0$ limit, corresponds to
$k_0=\iota'\,2k_F -\iota''\,k_F$ and reads,
\begin{equation}
q = k +\iota'\,2k_F -\iota''\,k_F\, ; \hspace{1cm} \omega_{c0,\,\iota',\,\iota''} (q) =
-\epsilon_{c0} (q) \, . \label{ciline}
\end{equation}
The $c0,\,+1,\,+1$ branch line, $c0,\,-1,\,-1$ branch line, and $c0,\,-1,\,+1$ branch
line are labeled $c$, $c'$, and $c''$ in Fig. 1 of Ref. \cite{EPL}, respectively, where
they are represented for $k>0$ and $\omega\leq 0$. In this case the value of the
functional (\ref{Delta}) is given by the $m\rightarrow 0$ limit of the following
parameter,
\begin{equation}
2\Delta_{\alpha\nu}^{\iota}(q) = \Bigl\{\iota'\,{\xi^1_{\alpha\nu\,c0}\over 2}
-\iota\,{\xi^0_{\alpha\nu\,s1}\over 2} - \iota''\,{\xi^1_{\alpha\nu\,s1}\over 2} -
\Phi_{\alpha\nu,\,c0}(\iota\,q^0_{F\alpha\nu},\,q)\Bigr\}^2 \, , \label{Delci}
\end{equation}
where $\alpha\nu = c0,\,s1$.

The dependence of this quantity on the momentum $k$ is obtained by combining Eqs.
(\ref{ciline}) and (\ref{Delci}). From use of the general expression (\ref{Ichiun}) in
the $m\rightarrow 0$ limit, we find the following expression for the one-electron removal
spectral function,
\begin{equation}
B (k,\,\omega) \approx C_{c0,\,\iota',\,\iota''} (q)\Bigl({-[\omega
+\omega_{c0,\,\iota',\,\iota''} (q)]\over
4\pi\sqrt{v_{c0}\,v_{s1}}}\Bigr)^{\zeta_{c0,\,\iota',\,\iota''} (q)} \, . \label{Bci}
\end{equation}
This expression corresponds to energy values just below the branch lines. In this
expression and in the exponent expressions provided below, the bare-momentum values are
in the range $-2k_F<q< 2k_F$. Furthermore, the corresponding momentum values belong to
the domains $-3k_F<k<k_F$ and $-k_F<k<3k_F$ for $\iota'=\iota''=1$ and
$\iota'=\iota''=-1$, respectively, and $-5k_F<k<-k_F$ and $k_F<k<5k_F$ for
$\iota'=-\iota''=1$ and $\iota'=-\iota''=-1$, respectively. In the $m\rightarrow 0$
limit, the exponent $\zeta_{c0,\,\iota',\,\iota''}(q)$ of expression (\ref{Bci}) reads,
\begin{eqnarray}
\zeta_{c0,\,\iota',\,\iota''}(q) & = & -1 + \sum_{\alpha\nu =c0,\,s1}\,\sum_{\iota =\pm
1}\, \Bigl\{ -\iota\iota'\,{\xi^1_{\alpha\nu\,c0}\over 2} + {\xi^0_{\alpha\nu\,s1}\over
2} + \iota\,\iota''\,{\xi^1_{\alpha\nu\,s1}\over 2} \nonumber \\
& + & \iota\,\Phi_{\alpha\nu,\,c0}(\iota\,q^0_{F\alpha\nu},\,q)\Bigr\}^2
\nonumber \\
& = & -1 + \sum_{\iota =\pm 1}\Bigl[\Bigl\{ -\iota\,{\xi_0\over
2}\Bigl(\iota'-{\iota''\over 2}\Bigr) + \iota\,\Phi_{c0,\,c0}(\iota\,2k_F,\,q)\Bigr\}^2 \nonumber \\
& + &
\Bigl\{{1\over\sqrt{2}}\Bigl(1+{\iota\,\iota''\over 2}\Bigr) +
\iota\,\Phi_{s1,\,c0}(\iota\,k_F,\,q)\Bigr\}^2\Bigr]  \, . \label{zetaci}
\end{eqnarray}
For $U/t>0$ the exponent $\zeta_{c0,\,+1,\,+1} (q)$ of Eq. (\ref{zetaci}) is negative for
all values of momentum, whereas $\zeta_{c0,\,-1,\,-1} (q)$ is also in general negative,
except for small values of $U/t$ and a small domain of bare-momentum values. These
exponents are plotted in Figs. 2 and 3, respectively, as a function of the momentum $k$
for $k>0$, several values of $U/t$, and electronic density $n=0.59$. In these figures
these exponents are called $\zeta_{c}$ and $\zeta_{c'}$, respectively. Correspondingly,
when $\zeta_{c0,\,\iota',\,\iota''}(q)<0$ the weight distribution (\ref{Bci}) describes a
singular branch line.

In turn, the exponents $\zeta_{c0,\,+1,\,-1}$ and $\zeta_{c0,\,-1,\,+1}$ of Eq.
(\ref{zetaci}) are positive. For the values of momentum for which these exponents are
smaller than one the spectral-function expression (\ref{Bci}) describes edge branch
lines. For finite values of $U/t$ the pre-factors $C_{c0,\,\pm 1,\,\mp 1} (q)$ have in
general smaller values than the pre-factors $C_{c0,\,\pm 1,\,\pm 1} (q)$. Moreover, for
finite values of $U/t$ the pre-factors $C_{c0,\,+1,\,+1} (q)$ and $C_{c0,\,-1,\,-1} (q)$
are decreasing and and increasing functions of $k$, respectively, whose values are
smallest for the domains $-3k_F<k<-2k_F$ and $2k_F<k<3k_F$, respectively. Again, in the
vicinity of the branch line end points $k=\pm k_F$, $k=\pm 3k_F$, and $k=\pm 5k_F$, where
$v_{c0}(q)\approx  \pm v_{c0}$, the spectral function is not of the form (\ref{Bci}), but
instead is of the general form given in Eq. (\ref{zeta*END}).

\begin{figure*}
\includegraphics[width=7cm,height=7cm]{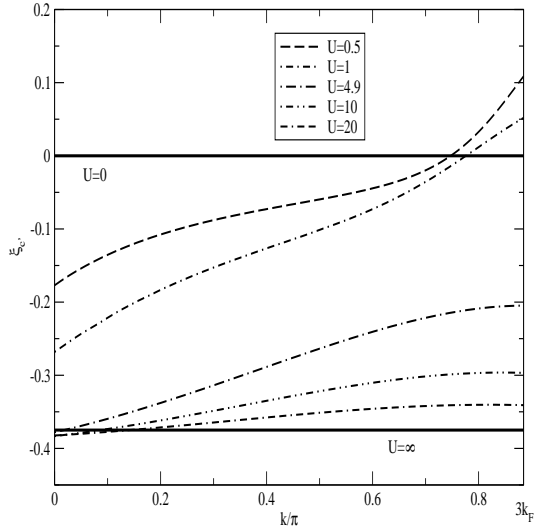}
\caption{\label{fig3} Momentum dependence of the exponent (\ref{zetaci}) along the
one-electron removal charge $c'\equiv c0,\,-1,\,-1$ branch line of Fig. 1 of Ref.
\cite{EPL} for $0<k<3k_F$. In the figure that exponent is called $\zeta_{c'}$ and is
given by $0$ and $-3/8$ for $U/t\rightarrow 0$ and $U/t\rightarrow\infty$, respectively.
For a small momentum domain in the vicinity of the branch-line end point $k=3k_F$, where
$v_{c0}(q)\approx v_{c0}$, this exponent does not apply, since the spectral function is
of the form given in Eq. (\ref{zeta*END}).}
\end{figure*}

The exponents plotted in Figs. 2 and 3 have different values in the limits $m\rightarrow
0$, $U/t\rightarrow 0$ and $U/t\rightarrow 0$, $m\rightarrow 0$. This justifies the
different values of these exponents given for $U/t\rightarrow 0$ in 
Figs. 2 and 3 and in Fig. 3 of Ref. \cite{EPL}, respectively. (Due to a misprint 
one of the $U/t$ dependent functions contributing to the 
expression used in Ref. \cite{EPL} for the exponent given in 
Eq. (\ref{zetaci}) was replaced by its large-$U/t$ asymptotic expansion; Therefore,
for small finite values of $U/t$ the values provided here for that exponent are also 
different from those given in that reference. Expression (\ref{zetaci}) 
and the corresponding Figs. 2 and 3 of this paper correct such a misprint. Fortunately, 
the only qualitative correction is that the exponent plotted in Fig. 3 becomes positive 
for small values of $U/t$ and a small domain of momentum values in the vicinity 
of $3k_F$.)

We finish this section by confirming that the momentum and energy dependence of the
spectral-weight distribution in the vicinity of the corresponding branch lines in the
limits $U/t\rightarrow 0$ and $U/t\rightarrow\infty$, recovers the correct behaviors. (We
recall that in this paper we reach the limit $U/t\rightarrow 0$ by considering the limit
$U/t\rightarrow 0$, $m\rightarrow 0$.) All expressions provided below are valid for
electronic densities $n$ such that $0<n<1$.

Independently of the general exponent expressions derived by the PDT of Ref. \cite{V-1},
we also used here the method of Refs. \cite{Penc96,Penc97} to derive the exponents
associated with the one-electron removal spectral function expressions obtained in these
references for $U/t\rightarrow\infty$. The limiting values of the exponents obtained here
fully agree with those obtained for $U/t\rightarrow\infty$ by use of the method of Refs.
\cite{Penc96,Penc97}. Thus, our general $U/t$ expressions are fully consistent
with the spectral function expressions found in these references for
$U/t\rightarrow\infty$.

Use of the general results of Ref. \cite{V-1} for the one-electron removal spectral
function in the vicinity of the $s1$ branch line, which for finite values of $U$ and
energy $\omega$ is given by expression (\ref{Bs}), reveals that as  $U/t\rightarrow 0$
the pre-factor $C_{s1}(q)$ corresponding to the general finite-energy expression
(\ref{Ichiun}) is replaced by the weight constant of the $\delta$-peak spectral function
given in the second expression of Eq. (\ref{Ichiun}). The regimen associated with such a
weight constant arises for very low values of $U/t$, where it is given by
$(1/N_a)^{\zeta_0 (q)}\rightarrow 1$ \cite{V-1}. Its independence on the value of the
bare-momentum $q$ results from the behavior of the functional $\zeta_0 (q)$, which for
this specific branch line is such that $\zeta_0 (q)\rightarrow 0$ as $U/t\rightarrow 0$
for the whole corresponding domain of $q$ values. We find that in the limits
$U/t\rightarrow 0$ and $U/t\rightarrow\infty$, the $s1$ branch-line exponent given in Eq.
(\ref{zetas}) reads,
\begin{equation}
\zeta_{s1}(q) = -1 \, , \hspace{0.25cm} U/t\rightarrow 0 \, ; \hspace{0.5cm}  \zeta_{s1}(q)
= - {1\over 2} + 2\Bigl({q\over 4k_F}\Bigr)^2 \, , \hspace{0.25cm} U/t\rightarrow\infty \,
, \label{limits}
\end{equation}
for the $q$ and $k$ values of the spectral-function expression (\ref{Bs}). In turn, the
pre-factor $C_{s1,\,\iota}$ corresponding to the low-energy expression (\ref{zeta*END})
is such that $C_{s1,\,\iota}\rightarrow 0$ as $U/t\rightarrow 0$. Thus, in that limit the
second expression of Eq. (\ref{Ichiun}) is valid for the whole branch-line bare-momentum
domain and the regimen associated with the spectral-function expression (\ref{zeta*END})
disappears. Moreover, according to Eq. (A2) of Ref. \cite{II}, in the limit
$U/t\rightarrow 0$ the dispersion $\epsilon_{s1}(q)$ becomes the electronic spectrum
$\epsilon_{s1}(q)=-2t[\cos (q)-\cos (k_F)]$. Consistently, according to Eq.
(\ref{limits}), the exponent (\ref{zetas}) is such that $\zeta_{s1}(q)\rightarrow -1$ as
$U/t\rightarrow 0$ for all values of $q$ in the range $0<\vert\, q\vert< k_F$ and thus of
the momentum $k$ in the domain $0<\vert\, k\vert< k_F$. Then, following the second
expression of Eq. (\ref{Bs}), the correct non-interacting one-electron removal spectral
function is reached in this limit.

For the one-electron removal $s1,\,\iota'$ branch line expression (\ref{Bsi}), the
multiplicative coefficient is such that $C_{s1,\,\iota'}(q)\rightarrow 0$ as
$U/t\rightarrow 0$. Thus, such a branch line does not exist for $U/t\rightarrow 0$, which
is the correct result. For $m\rightarrow 0$ and in the limits $U/t\rightarrow 0$ and
$U/t\rightarrow\infty$ the corresponding exponent (\ref{zetasi}) reads,
\begin{equation}
\zeta_{s1,\,\iota'}(q) = 3 \, , \hspace{0.25cm} U/t\rightarrow 0 \, ; \hspace{0.5cm}
\zeta_{s1,\,\iota'}(q) = {3\over 2} -\iota'\,{q\over k_F} + 2\Bigl({q\over 4k_F}\Bigr)^2
\, , \hspace{0.25cm} U/t\rightarrow\infty \, , \label{limitsi}
\end{equation}
for the $q$ and $k$ values of the spectral-function expression (\ref{Bsi}). In the limit
$U/t\rightarrow\infty$, this exponent is such that $\zeta_{s1,\,\iota'}(q) = 5/8$ for
$q\rightarrow \iota'\,k_F$, $\zeta_{s1,\,\iota'}(q) = 3/2$ for $q = 0$, and
$\zeta_{s1,\,\iota'}(q) = 21/8$ for $q\rightarrow -\iota'\,k_F$. For the one-electron
removal $c0,\,\iota',\,\iota''$ branch line expression (\ref{Bci}), the multiplicative
coefficient is such that $C_{c0,\,\iota',\,\iota''}(q)\rightarrow 0$ as $U/t\rightarrow
0$, which again is the correct result. In the limit $U/t\rightarrow 0$, the
$c0,\iota',\iota''$ branch lines disappear, all spectral weight being transferred over to
the $s1$ branch line, which becomes the non-interacting one-electron removal spectrum. As
the limits $U/t\rightarrow 0$ and $U/t\rightarrow\infty$ are approached, the exponent
(\ref{zetaci}) tends to the following values,
\begin{eqnarray}
\zeta_{c0,\,\iota',\,\iota''}(q) & = & {(\iota'-\iota'')^2\over 2} \, ; \hspace{0.15cm}
U/t\rightarrow 0 \, ; \nonumber \\
\zeta_{c0,\,\iota',\,\iota''}(q) & = & -{\iota'\,\iota''\over 2} + {1\over 8} \, ; \hspace{0.15cm} U/t\rightarrow\infty \, ,
\label{limitscsiU0}
\end{eqnarray}
for the $q$ and $k$ values of the spectral-function expression (\ref{Bci}). Thus, in the
limit $U/t\rightarrow 0$, it is given by $0$ for the branch lines such that
$\iota'\,\iota''=1$ and $2$ for the branch lines such that $\iota'\,\iota''=-1$.
Furthermore, for $U/t\rightarrow\infty$ the exponent is given by $-3/8$ for the branch
lines such that $\iota'\,\iota''=1$ and $5/8$ for the branch lines such that
$\iota'\,\iota''=-1$.

Hence, the one-electron removal $s1$ branch line becomes the non-interacting removal
electronic spectrum, which corresponds to $-k_F< k< k_F$. In turn, for finite values of
$U/t$ the spectral weight spreads over a larger two-dimensional region of the
$(k,\,\omega)$-plane. However, most of the spectral weight is located in the vicinity of
separated and independent $c0$ and $s1$ branch lines and of the weak border line
mentioned in Sec. IV. Our study provides the momentum and energy dependence of the weight
distribution in the vicinity of such $\alpha\nu$ branch lines. In the $m\rightarrow 0$
limit, the maximum spread of the one-electron spectral-weight distribution occurs for
$U/t\rightarrow\infty$, where the problem had been already studied in Refs.
\cite{Penc96,Penc97}. The $U/t\rightarrow\infty$ maximum spreading of the
one-electron removal spectral weight at electronic density $n=1/2$ is illustrated in Fig.
1 of Ref. \cite{Penc96} for the spectral function $B (k,\,\omega)$. Our
$U/t\rightarrow\infty$ expressions of Eqs. (\ref{limits}) and (\ref{limitscsiU0}) agree
with the results obtained by the method of Refs. \cite{Penc96,Penc97}, as
mentioned above.

\section{DISCUSSION ABOUT THE RELATION TO THE PHOTOEMISSION DISPERSIONS OF
TTF-TCNQn AND CONCLUDING REMARKS}

An interesting realization of a quasi-1D metal is the organic charge-transfer salt
TTF-TCNQ \cite{Basista,spectral0,Ralph}. The experimental dispersions in the electron
removal spectrum of this quasi-1D conductor as measured by ARPES are shown in Fig. 9 (b)
of Ref. \cite{spectral0} and Fig. 4 of Ref. \cite{EPL}. The experimental data in these
figures were taken with He I radiation (21.2 eV) at a sample temperature of 60 K on a
clean surface obtained by {\it in situ} cleavage of a single crystal. Instrumental energy
and momentum resolution amounted to 70 meV and 0.07 \AA$^{-1}$, respectively.

We note that the low-energy spectral properties of TTF-TCNQ involve inter-chain hopping
and electron-phonon interactions. Thus, the 1D Hubbard model PDT results are to be
applied above the energies of these processes. The singular branch lines studied in Sec.
III correspond to the $(k,\,\omega)$-plane region which contains all one-electron removal
spectral-weight singular features. In spite of the recent improvements in the resolution
of photoemission experiments \cite{spectral0,Ralph,Zwick}, it is difficult to measure the
exponents and the finest details of the electronic structure experimentally, in part due
to the extrinsic losses that occur on very anisotropic conducting solids \cite{Joynt}.
Thus, a crucial test for the suitability of the model (\ref{H}) to describe real quasi-1D
materials is whether the ARPES peak dispersions correspond to the singular branch lines
and other divergent spectral features predicted by the PDT of Ref. \cite{V-1}.

The electronic density of TCNQ is $n=0.59<1$. For densities in the domain $0<n<1$ and
one-electron removal, the main singular spectral features predicted by the general PDT
are of branch line type. Thus, for TCNQ the main divergent spectral features correspond
to the singular branch lines studied in Sec. III. The only other singular feature is
quite weak and corresponds to the lowest line of Fig. 1 of Ref. \cite{EPL}. In the
vicinity of such border line the spectral-weight distribution corresponds to the
power-law dependence (\ref{B-border}), which is controlled by a $U/t$ independent
exponent. Due to its weakness such a border line does not lead to any preeminent TCNQ
spectral feature.

While the theoretical weight-distribution branch-line expressions provided in Sec. III
refer to all values of $U/t$ and $n$, a detailed study of the spectral-function $k$,
$\omega$, and $U/t$ dependence in the vicinity of the branch lines obtained in this paper
confirms the validity of the preliminary predictions of Refs. \cite{spectral0,EPL}: the
electron removal spectra calculated for $t = 0.4$ eV, $U = 1.96$ eV ($U/t = 4.90$), and
$n=0.59$ yields an almost perfect agreement with the three TCNQ experimental dispersions.
The exception is the low-energy behavior, as a result of the inter-chain hopping and
electron-phonon interactions, as mentioned above. If accounted for a renormalization of
the transfer integral due to a possible surface relaxation \cite{spectral0}, these values
are in good agreement with estimates from other experiments \cite{Zwick,Kagoshima}.

The experimental TCNQ finite-energy peak dispersions of Fig. 4 of Ref. \cite{EPL}
correspond to the spin $s\equiv s1$ branch line (\ref{sline}) and charge $c\equiv
c0,\,+1,\,+1$ and $c'\equiv c0,\,-1,\,-1$ branch lines (\ref{ciline}) of Fig. 1 of that
reference. Those are the main finite-weight singular branch lines in the one-electron
removal spectral function for $U/t = 4.90$ and $n=0.59$. Importantly, only these main
singular features, whose line shape is controlled by negative exponents, lead to TCNQ
peak dispersions in the real experiment. The exponent (\ref{zetas}) corresponds to the
spin $s\equiv s1$ branch line and is plotted in Fig. 1 for $0<k<k_F$. The exponents
(\ref{zetaci}) that correspond to the charge $c\equiv c0,\,+1,\,+1$ branch line and
charge $c'\equiv c0,\,-1,\,-1$ branch line are plotted in Figs. 2 and 3, respectively. As
reported in Sec. III, for finite values of $U/t$ the value of the constant
$C_{c0,\,-1,\,-1}(q)$ of the spectral-function expression (\ref{Bci}) strongly decreases
for momentum values such that $2k_F<k<3k_F$. This is consistent with the absence of TCNQ
experimental spectral features for momentum values $k>0.59\pi\approx 0.50$ \AA$^{-1}$ in
Fig. 4 of Ref. \cite{EPL}, along the corresponding $c'\equiv c0,\,-1,\,-1$ branch line of
Fig. 1 of that reference.

Thus, our detailed branch-line PDT analysis fully agrees with the preliminary theoretical
results of Refs. \cite{spectral0,EPL} for the TCNQ problem. On the other hand, the
theoretical predictions for the TTF dispersions presented in Ref. \cite{EPL} are very
preliminary. For the electronic density value corresponding to the TCNQ stacks the main
singular spectral features are of branch-line type and the only existing border line is
quite weak. In contrast, a careful analysis of the problem by means of the general PDT
reveals that for the electronic density suitable to the TTF stacks the main singular
features are both of branch-line and border-line type. Once the preliminary studies of
TTF presented in Ref. \cite{EPL} involve the singular branch line features only, a very
small value of $U/t$ is predicted. However, if instead one takes into account all
singular features provided by the PDT, the best quantitative agreement with the TTF
experimental dispersions is reached for larger values of $U/t$, as confirmed elsewhere.

In this paper we have used the exact PDT of Ref. \cite{V-1} to study the energy and
momentum dependence of the one-electron removal spectral weight distribution in the
vicinity of the singular and edge branch lines of the 1D Hubbard model. A careful and
detailed analysis of the spectral function expressions in the proximity of the charge and
spin branch lines obtained here confirms the validity of the preliminary theoretical
predictions of Refs. \cite{spectral0,EPL}, in what the description of the band TCNQ
dispersions observed by ARPES in the quasi-1D organic compound TTF-TCNQ is concerned. The
TCNQ conduction band displays spectroscopic signatures of spin-charge separation on an
energy scale of the band width. This seems to indicate that the dominant non-perturbative
many-electron microscopic processes studied in Ref. \cite{V-1} by means of the PDT and
the associated scattering mechanisms investigated in Ref. \cite{S0} control the unusual
finite-energy spectral properties of TTF-TCNQ. The quantitative agreement for the whole
finite-energy band width between the 1D Hubbard model PDT theoretically predicted
spectral features and the TCNQ photoemission dispersions of TTF-TCNQ reveals that for
finite-energy the local effects of the Coulomb electronic correlations fully control the
spectral properties of that material. Thus, we expect that the long-range Coulomb
interactions, disorder, and impurity effects play very little role in the finite-energy
and/or finite-temperature properties of TTF-TCNQ. That disorder and impurities do not
play a major role is confirmed by the occurrence of spin-charge separation for the whole
energy band width. Indeed, the presence of disorder and impurities would prevent the
separation of the one-electron spectral-weight distribution in terms of spin and charge
singular spectral features.

Our present finite-energy description goes beyond the usual TLL low-energy investigations
by means of bosonization \cite{Schulz} and conformal-field theory \cite{CFT}. For low
energy the present quantum problem is a TLL. This concept only applies to the parts of
the one-electron spectrum of Fig. 1 of Ref. \cite{EPL} where the spectral dispersions can
be linearized. From analysis of the figure branch lines one finds that such a regimen
corresponds to low energies. However, our results refer to all values of the group
velocities associated with the branch lines plotted in that figure. Thus, the spin-charge
separation found here corresponds to the whole finite-energy band width. Only our
finite-energy theoretical spectral features describe the experimental photoemission TCNQ
dispersions of TTF-TCNQ, once the low-energy phase of TTF-TCNQ is not metallic and
corresponds instead to a broken-symmetry state \cite{Basista}. It follows that for the
present TCNQ photoemission problem, the 1D physics described by the 1D Hubbard model only
becomes experimentally relevant for finite energy, where the low-energy TLL description
does not apply.

A detailed theoretical study of the the TTF experimental dispersions by means of the PDT,
including consideration of both singular branch lines studied here and singular border
lines is in progress and will be presented elsewhere. Moreover, the calculation of the
one-electron spectral-function of the 1D Hubbard model for all values of $k$ and $\omega$
by use of the general PDT, which consider all contributing processes, is also in
progress.

{\bf Acknowledgments} \vspace{0.2cm}

We thank A. Bjelis, D. Bozi, A. Castro Neto, F. Guinea, E. Jeckelmann, P. A. Lee, J. M.
B. Lopes dos Santos, L. M. Martelo, J. P. Pouget, and U. Schwingenschl\"ogl for
discussions. J.M.P.C., K.P., and P.D.S. thank the support of the ESF Science Programme
INSTANS 2005-2010, J.M.P.C. and P.D.S. that of the FCT grant POCTI/FIS/58133/2004,
J.M.P.C. the hospitality and support of MIT where part of this research was
fulfilled and the support of the Calouste Gulbenkian Foundation and
Fulbright Commission, K.P. that of the OTKA grant T049607, and R.C. thanks 
the support of Deutsche Forschungsgemeinschaft (CL 124/3-3).




\section*{References}
\begin{thebibliography}{100}
\bibitem[1]{Kagoshima}
        S. Kagoshima, H. Nagasawa, T. Sambongi, {\em
        One-dimensional conductors} (Springer, Berlin, 1987), and references
        therein.
\bibitem[2]{Basista}
        Basista H, Bonn DA, Timusk T, Voit J, J\'erome D and
        Bechgaard K 1990 {\it Phys. Rev. B} {\bf 42}, 4088.
\bibitem[3]{BS}
        Carmelo JMP, Horsch P,  Campbell DK and Castro Neto AH 1993 {\it Phys. Rev. B (RC)}
        {\bf 48}, 4200.
\bibitem[4]{spectral0}
        Sing M, Schwingenschl\"ogl U, Claessen R, Blaha P, Carmelo JMP,
        Martelo LM, Sacramento PD, Dressel M and Jacobsen CS 2003 {\it Phys. Rev. B} {\bf 68}, 125111.
\bibitem[5]{Ralph}
        Claessen R, Sing M, Schwingenschl\"ogl U, Blaha P,
        Dressel M, Jacobsen CS 2002 {\it Phys. Rev. Lett.} {\bf 88}, 096402.
\bibitem[6]{Zwick}
        Zwick F, J\'erome D, Margaritondo G, Onellion M, Voit J,
        Grioni M 1998 {\it Phys. Rev. Lett.} {\bf 81}, 2974.
\bibitem[7]{Lieb}
        Lieb Elliott H and Wu FY 1968 {\it Phys. Rev. Lett.} {\bf 20},
        1445.
\bibitem[8]{Takahashi}
        Takahashi M 1972 {\it Prog. Theor. Phys.} {\bf 47}, 69.
\bibitem[9]{properties}
        Peres NMR, Carmelo JMP, Campbell DK and Sandvik AW 1997
        {\it Zeitsch. Phys. B} {\bf 103}, 217; Baeriswyl D, Carmelo J and Maki K 1987
        {\it Synth. Met.} {\bf 21}, 271.
\bibitem[10]{Lee}
        Simons BD, Lee PA and Altshuler BL 1993
        {\it Phys. Rev. Lett.} {\bf 70}, 4122; Arikawa M, Saiga Y and 
        Kuramoto Y 2001 {\it Phys. Rev. Lett.} {\bf 86}, 3096
        (2001);Penc K and Shastry BS, 2002 {\it Phys. Rev. B} 
        {\bf 65}, 155110.
\bibitem[11]{Schulz}
        Schulz H J 1990 {\it Phys. Rev. Lett.} {\bf 64}, 2831;
        Carmelo JMP, Castro Neto AH and Campbell DK 1994 {\it Phys. Rev. Lett.} {\bf 73}, 926
        and Erratum 1995 {\bf 74}, 3089; Carmelo JMP, Castro Neto AH and Campbell DK 1994
        {\it Phys. Rev. B} {\bf 50}, 3683.
\bibitem[12]{Woy}
        Woynarovivh F 1989 {\it J. Phys. A} {\bf 22}, 4243.
\bibitem[13]{Ogata}
        Ogata M and Shiba H 1990 {\it Phys. Rev. B} {\bf 41}, 2326.
\bibitem[14]{Kawakami}
        Kawakami N and Yang SK 1990 {\it Phys. Lett. A} {\bf 148}, 359.
\bibitem[15]{Frahm}
        Frahm H and Korepin VE 1990 {\it Phys. Rev. B} {\bf 42}, 10 553.
\bibitem[16]{Brech}
        Brech M, Voit J and Buttner H 1990 {\it Europhys. Lett.} {\bf 12}, 289.
\bibitem[17]{CFT}
         Frahm H and Korepin VE 1991 {\it Phys. Rev. B} {\bf 43}, 5653.
\bibitem[18]{Ogata91}
        Ogata M, Sugiyama T and Shiba H  1991 {\it Phys. Rev. B} {\bf 43}, 8401.
\bibitem[19]{Karlo}
        Penc K and S\'olyom J 1993 {\it Phys. Rev. B} {\bf 47}, 6273 (1993).
\bibitem[20]{algebras}
        Carmelo JMP and Castro Neto AH 1993 {\it Phys. Rev. Lett.} {\bf 70}, 1904;
        Carmelo JMP, Castro Neto AH and Campbell DK 1994
        {\it Phys. Rev. B} {\bf 50}, 3667 (1994).
\bibitem[22]{Penc96}
        Penc K, Hallberg K, Mila F and Shiba H 1996 {\it Phys. Rev. Lett.}
        {\bf 77}, 1390.
\bibitem[23]{Penc97}
        Penc K, Hallberg K, Mila F and Shiba H 1997 {\it Phys. Rev. B} {\bf 55}, 15 475.
\bibitem[24]{Sorella}
        Sorella S and Parola A 1996 {\it Phys. Rev. Lett.} {\bf 76}, 4604 (1996).
\bibitem[25]{Senechal}
        S\'en\'echal D, Perez D and Pioro-Ladri\`ere M 2000 {\it Phys. Rev. Lett.} {\bf 84}, 522 (2000).
\bibitem[26]{I}
        Carmelo JMP, Rom\'an JM, and Penc K 2004 {\it Nucl. Phys. B} {\bf 683}, 387.
\bibitem[27]{V-1}
        Carmelo JMP, Penc K and Bozi D 2005 {\it Nucl. Phys. B} {\bf 725},
        421; Carmelo JMP and Penc K, to appear in the {\it Europ. Phys.
        J. B} (cond-mat/0311075).
\bibitem[28]{S0}
        Carmelo JMP 2005 {\it J. Phys.: Cond. Mat.} {\bf 17}, 5517.
\bibitem[29]{EPL}
        Carmelo JMPC, Penc K, Martelo LM, Sacramento PD,
        Lopes dos Santos JMB, Claessen R, Sing M and
        Schwingenschl\"ogl U 2004 {\it Europhys. Lett.} {\bf 67}, 233.
\bibitem[30]{Eric}
        Benthien H, Gebhard F and Jeckelmann E 2004 {\it Phys. Rev. Lett.}
        {\bf 92}, 256401.
\bibitem[31]{LE}
        Carmelo JMP and Penc K, cond-mat/0508704; Carmelo JMP,
        Martelo LM and Penc K 2006 at press in {\it Nucl. Phys. B}.
\bibitem[32]{super}
        Carmelo JMP, Guinea F, Penc K and Sacramento PD 2004 {\it Europhys. Lett.} {\bf 68}, 839.
\bibitem[33]{tj}
        Bares PA, Carmelo JMP, Ferrer J and Horsch P 1992 {\it Phys. Rev. B} {\bf 46} 14 624.
\bibitem[34]{II}
        Carmelo JMP and Sacramento PD 2003 {\it Phys. Rev. B} {\bf 68}, 085104.
\bibitem[35]{Joynt}
        Joynt R 2000 {\it Science} {\bf 284}, 777.
\end{thebibliography}
\end{document}